\documentclass[]{aastex631}

\usepackage{amsmath}
\shorttitle{Equilibrium Q}
\shortauthors{Forbes}
\graphicspath{{./}{figures/}}

\begin{document}

\title{How Low Can Q Go?}

\author[0000-0002-1975-4449]{John C. Forbes}
\affiliation{Center for Computational Astrophysics 
Flatiron Institute 
162 5th Avenue 
New York, NY, 10010, USA}
\correspondingauthor{John C. Forbes}
\email{johncforbes@gmail.com}

\begin{abstract}

Gravitational instability plays a substantial role in the evolution of galaxies. Various schemes to include it in galaxy evolution models exist, generally assuming that the Toomre $Q$ parameter is self-regulated to $Q_\mathrm{crit}$, the critical $Q$ dividing stable from unstable conditions in a linear stability analysis. This assumption is in tension with observational estimates of $Q$ that find values far below any plausible value of $Q_\mathrm{crit}$. While the observations are subject to some uncertainty, this tension can more easily be relieved on the theoretical side by relaxing the common assumption that $Q\ge Q_\mathrm{crit}$. Based on observations of both $z\sim 2$ disks and local face-on galaxies, we estimate the effect of gravitational instability necessary to balance out every other physical process that affects $Q$. In particular we find that the disk's response to low $Q$ values can be described by simple functions that depend only on $Q$. These response functions allow galaxies to maintain $Q$ values below $Q_\mathrm{crit}$ in equilibrium over a wide range of parameters. Extremely low values of $Q$ are predicted when the gas surface density is $\ga 10^3 M_\odot\ \mathrm{pc}^{-2}$, the rotation curve provides minimal shear, the orbital time becomes long, and/or when the gas is much more unstable than the stellar component. We suggest that these response functions should be used in place of the $Q\ge Q_\mathrm{crit}$ ansatz.

\end{abstract}

\section{Introduction} \label{sec:intro}

A variety of theoretical and observational results over the past two-plus decades have suggested that gravitational instability, fundamentally like the linear instability in axisymmetric self-gravitating disks discovered by \citet{toomre_gravitational_1964}, may play a substantial role in the evolution of galaxies. $Q$ summarizes the stability of such disks, where $Q$ below some critical value $Q_\mathrm{crit}$ of order 1 indicates that the disk will be subject to clumping or the formation of spiral arms. In local galaxies where surface densities and kinematics of both gaseous and stellar components of galaxies may be measured, the Toomre $Q$ parameter is observed to be in the neighborhood of 2 across many different galaxies \citep{leroy_star_2008}. At $z\sim 2$, gas velocity dispersions \citep[e.g.][]{kassin_epoch_2012, genzel_sins_2014, wisnioski_kmos3d_2015, forsterschreiber_sins_2018,ubler_evolution_2019,price_mosdef_2020} far in excess of the level likely to be sustainable by supernova feedback alone \citep{joung_dependence_2009, krumholz_unified_2018, brucy_largescale_2020}, along with clumpy UV morphologies \citep{elmegreen_resolved_2007} have also suggested  that gravitational instability had a substantial role to play. Indeed \citet{genzel_sins_2014} has reported estimates of the Toomre $Q$ well below $Q_\mathrm{crit}$. There are also a number of local examples of $Q$ falling below $Q_\mathrm{crit}$ \citep{westfall_diskmass_2014} from the DiskMass Survey \citep{bershady_diskmass_2010}.

In modeling these disks it has been convenient to assume that $Q=Q_\mathrm{crit}$ \citep[e.g.][]{thompson_radiation_2005, dekel_formation_2009, krumholz_dynamics_2010, forbes_evolving_2012, faucher-giguere_feedbackregulated_2013, krumholz_turbulence_2016, hayward_how_2017, krumholz_unified_2018}, or at least that $Q$ may not fall below $Q_\mathrm{crit}$ \citep{forbes_balance_2014, rathaus_stellar_2016, forbes_radially_2019}. Physically the idea is that when $Q$ falls below $Q_\mathrm{crit}$, the resulting fragmentation, formation of clumps, star formation, and/or torques exerted by the resulting inhomogeneous gravitational potential, will tend to increase the velocity dispersion and drive $Q$ back to $Q_\mathrm{crit}$. This assumption has enabled, under various additional assumptions, the calculation of star formation rates, mass loading factors, mass transport rates, viscous evolution timescales, and contributions to the turbulent energy of the disk. This process does occur in simulations, where many groups have found that galactic disks tend to self-regulate to $Q \sim 1$ \citep{ bournaud_unstable_2009, goldbaum_mass_2015, goldbaum_mass_2016, behrendt_clusters_2016, gurvichPressureBalanceMultiphase2020} in an azimuthally-averaged sense.

The exact value of $Q_\mathrm{crit}$ is complicated by several factors. First, galactic disks have multiple components (gas and stars), and indeed more if one requires that each component be describable by a single temperature or velocity dispersion. Each quasi-isothermal component is described by its own $Q$ parameter,
\begin{equation}
\label{eq:qi}
    Q_i = \frac{\kappa \sigma_i}{\pi G \Sigma_i}
\end{equation}
where $\kappa=\sqrt{2(\beta+1)}\Omega$ is the epicyclic frequency, $\Omega=v_\mathrm{circ}/r$ is the orbital frequency of an object orbiting on a circular orbit at velocity $v_\mathrm{circ}$, and $\beta = d\ln v_\mathrm{circ}/d\ln r$ is the logarithmic derivative of $v_\mathrm{circ}$. The velocity dispersion and column density of the $i$th component are $\sigma_i$ and $\Sigma_i$ respectively. For a single-component disk, equation \ref{eq:qi} is exactly $Q$. The gravitational stability of a gas-star mixture is approximately $Q^{-1} \sim Q_g^{-1} + Q_*^{-1}$ \citep{wang_gravitational_1994}. Intuitively the most unstable component dominates the stability of the mixture. More accurate approximations \citep{romeo_effective_2011} account for the finite thickness of the disk and deviations from the particularly simple \citet{wang_gravitational_1994} approximation. With more components, the most unstable mode (which sets $Q$) can be found numerically as well \citep{rafikov_local_2001}. Closer examination of the dispersion relation from which the $Q$ stability thresholds are derived has yielded the realization that the presence of fast cooling in galactic disks means that $Q$ needs to be substantially higher than the formal linear instability threshold, i.e  $Q_\mathrm{crit}$ should be closer to 2 or 3 than 1, to realistically avoid Toomre-like instabilities \citep{elmegreen_gravitational_2011}. Perhaps because finite thickness and multi-component corrections are often difficult to estimate directly from data, especially outside the local Universe, and perhaps because even the gas-only measurements are challenging with a wide variety of systematic uncertainties \citep{davies_how_2011}, little concern has been devoted to the discrepancy between $Q_\mathrm{crit} \approx 2-3$, and the observed $Q_\mathrm{gas} \approx 0.2$ in $z\sim 2$ surveys.

In this work we relax the $Q\ge Q_\mathrm{crit}$ assumption and derive equilibrium $Q$ values. Importantly these equilibrium values of $Q$ are not about what $Q_\mathrm{crit}$ itself is, but rather how far below $Q_\mathrm{crit}$ $Q$ may go. In the next section we show the key governing equations and discuss the uncertainties in this formulation, then in section \ref{sec:data} we show how these uncertainties may be at least partially resolved by estimating how quickly $Q$ is changing from everything besides gravitational instability in the observed galaxies. The results are interpreted in section \ref{sec:interpretation} and the implications are discussed in section \ref{sec:implications}. We conclude in section \ref{sec:conclusion}.

\section{Governing Equations}

Our goal is to find $Q_\mathrm{eq}$, the value of $Q$ at which 
\begin{equation}
\label{eq:eq1}
    \frac{dQ}{dt} = 0,
\end{equation}
in the regime where gravitational instability is active, i.e. $Q \le Q_\mathrm{crit}$. Equation 
\ref{eq:eq1} may be re-expressed using the chain rule, and by dividing up the resulting terms into those arising from gravitational instability, which we denote ``GI,'' and all other terms, which we call ``sources.''
\begin{equation}
\label{eq:2}
    \frac{dQ}{dt} = 0 = \frac{dQ}{dt}\Big|_\mathrm{GI} +  \frac{dQ}{dt}\Big|_\mathrm{sources}  = \frac{\partial Q}{\partial \Sigma}\frac{\partial \Sigma}{\partial t} + \frac{\partial Q}{\partial \sigma}\frac{\partial \sigma}{\partial t} + \frac{\partial Q}{\partial \Sigma_*}\frac{\partial \Sigma_*}{\partial t} + \frac{\partial Q}{\partial \sigma_*}\frac{\partial \sigma_*}{\partial t}
\end{equation}
In principle more terms can be added to the chain rule expansion, e.g. derivatives with respect to other components of $Q$ like $v_\mathrm{circ}$, but we assume that these terms are small \citep{krumholz_dynamics_2010, forbes_evolving_2012}. The main issue in evaluating the terms in Equation \ref{eq:2} is deciding which formulation of $Q$ itself to use. It is plausible that only $Q_\mathrm{gas}$ matters, especially in the gas-rich $z\sim 2$ disks on which we are focused, in which case the terms with $\Sigma_*$ and $\sigma_*$ would be approximately zero. Unless otherwise noted, we will adopt the \citet{romeo_effective_2011} approximation to $Q$ since it accounts for the two components we consider here and for finite thickness effects. This approximation has the disadvantage of involving a conditional depending on which component of the disk contributes more to $Q$, so we will carry through the partial derivatives with respect to $Q$ without explicitly evaluating them analytically.

\subsection{Source Terms}
\label{sec:sources}

The source terms come from changes in the state of the disk that are unrelated to the direct effects of gravitational instability. We enumerate these terms as follows, recalling that the quantities in parentheses following each $\partial Q/\partial X$ are just $\partial X/\partial t$ arising from any physical process besides gravitational instability (see Eq. \ref{eq:2}).  
\begin{align} \label{eq:sources}
    \frac{dQ}{dt}\Big|_\mathrm{sources} =& \frac{\partial Q}{\partial \Sigma}\left( \dot{\Sigma}_\mathrm{accr} - (f_R+\eta)\dot{\Sigma}_\mathrm{SF} \right) + \\ \nonumber
    & \frac{\partial Q}{\partial \sigma}\left( \frac{\langle p_*/m_* \rangle \dot{\Sigma}_\mathrm{SF}}{3\Sigma} -  \frac{\sigma^2}{2 H}\left(1 - \frac{\sigma_\mathrm{therm}^2}{\sigma^2} \right)^{3/2} + \frac{\epsilon_\mathrm{accr} \dot{\Sigma}_\mathrm{accr} v_\mathrm{circ}^2}{3 \Sigma \sigma} \right) + \\ \nonumber
    & \frac{\partial Q}{\partial \Sigma_*} \left(f_R \dot{\Sigma}_\mathrm{SF}\right) +\\ \nonumber
    & \frac{\partial Q}{\partial \sigma_*} \left( f_R \dot{\Sigma}_\mathrm{SF} \frac{\sigma^2 - \sigma_*^2}{2 \Sigma_* \sigma_*} \right) + \\ \nonumber
    & \frac{\partial Q}{\partial Q_*} \left( \frac{Q_*-Q_{*,\mathrm{crit}}}{\gamma t_\mathrm{orb}} \mathcal{I}_{Q_*} \right).
\end{align}
The change in surface density per unit time at any point in the disk is the sum of the effects of accretion, $\dot{\Sigma}_\mathrm{accr}$, and star formation plus outflows, $(f_R+\eta)\dot{\Sigma}_\mathrm{SF}$. Accretion is assumed to be set externally, e.g. by the physics of cosmological accretion \citep[e.g.][]{keres_how_2005,dekel_cold_2009}, although in many galaxies it may be more reasonable to tie it to past, even recent star formation via the action of galactic fountains or star-formation induced condensation \citep{oppenheimer_feedback_2010, hobbs_growing_2015}. The effect of star formation is reduced by a fraction $f_R \approx 0.5$ the fraction of material that remains in long-lived stars or stellar remnants rather than being returned to the ISM through stellar winds and supernovae \citep{tinsley_evolution_1980, leitner_fuel_2011}. Star formation is also assumed to remove gas from the galaxy in galactic winds with mass loading factor $\eta$. In principle we could have included a term proportional to $(1/2\pi r)(\partial \dot{M}/\partial r)$, the change in surface density arising from a mass flow through the disk $\dot{M}$ when that flow rate varies with radius. We assume, however, that $\dot{M}$ is primarily set by gravitational instability in this regime where $Q\le Q_\mathrm{crit}$, so this term is not included among the source terms $dQ/dt|_\mathrm{sources}$. 

The velocity dispersion of the gas, similarly, may be affected by gravitational stability, but that effect is not included among the source terms in Eq. \ref{eq:sources}. Instead, we include the effects from star formation feedback (first term), turbulent dissipation (second term), and direct accretion (third term). The star formation feedback term includes the factor $\langle p_*/m_*\rangle$, the amount of momentum injected by supernovae per unit mass formed in stars. Generally simulations find values of around $3000\ \mathrm{km}\ \mathrm{s}^{-1}$ \citep[e.g.][]{kim_momentum_2015}, but for sufficiently high gas fractions and long dynamical times, \citet{orr_bursting_2022} find that this number can be reduced substantially by the breakout of superbubbles from the disk before they finish their energy-driven phase. We adopt the \citet{orr_bursting_2022} values of $\langle p_* /m_* \rangle$ unless otherwise noted\footnote{In particular we evaluate their equation 8 for the velocity of the superbubble at $t=t_\mathrm{SNe}$, the time that supernovae continue to go off after a burst of star formation. If this velocity is greater than the local velocity dispersion, we evaluate their Equation 19 for the momentum per unit mass of stars formed. We do not adopt their simplifications like assuming that $Q_\mathrm{gas}$ is 1.}. The dissipation term assumes that turbulent energy dissipates on a scaleheight crossing time \citep{stoneDissipationCompressibleMagnetohydrodynamic1998, maclowKineticEnergyDecay1998}, and as in \citet{forbes_balance_2014} this dissipation is assumed to gradually approach zero as $\sigma \rightarrow \sigma_\mathrm{therm} = 8\ \mathrm{km}\ \mathrm{s}^{-1}$, the sound speed of the warm neutral medium. We adopt the expression for $H$ from \citet{ostriker_pressureregulated_2022} (see below). For the contribution from direct accretion, we adopt the formula suggested by \citet{klessen_accretiondriven_2010} with $\epsilon_\mathrm{accr} = 10\%$ as measured in Illustris TNG50 by \citet{forbes_gas_2022}.

The remaining three terms summarize the effect of changes to the stars, namely changes to $\Sigma_*$ and $\sigma_*$, on $Q$. Just as star formation removes gas, it adds to $\Sigma_*$. The stellar population as a whole also loses specific kinetic energy by the addition of new stars that have the same velocity dispersion as the (dynamically colder) gas. Stars may also heat up via their interactions with spiral arms \citep{sellwood_spiral_1984, carlberg_dynamical_1985} and other structures in the disk like molecular clouds \citep{lacey_influence_1984}. We assume that these effects follow the \citet{sellwood_spiral_1984} and \citet{carlberg_dynamical_1985} prescription that $Q_*$ increases in proportion to its difference from $Q_{*,\mathrm{crit}}$ per unit orbital time, so long as $Q_*<Q_{*,\mathrm{crit}}$. This last requirement is encapsulated in the indicator function $I_{Q_*}$, which is 1 when $Q_*<Q_{*,\mathrm{crit}}$ and zero otherwise. We adopt $\gamma=4$ and $Q_{*,\mathrm{crit}}=2$. Note that this rate of change in $Q_*$ due to spiral heating could be decomposed into its effects on $\Sigma_*$ and $\sigma_*$ \citep[e.g.][]{forbes_balance_2014}, but for our purposes here all we care about is its effects on $Q$, so such a decomposition is not necessary.

We also need to specify exactly how we estimate $H$, the star formation rate $\dot{\Sigma}_\mathrm{SF}$, and the accretion rate $\dot{\Sigma}_\mathrm{accr}$. Following \citet{ostriker_pressureregulated_2022}, we adopt
\begin{equation}
\label{eq:H}
H = \frac{2 \sigma^2}{\pi G \Sigma + ((\pi G\Sigma)^2 + 32\pi \zeta_d G \rho_\mathrm{sd} \sigma^2)^{1/2}},
\end{equation}
where $\zeta_d\approx 0.33$ depending on the exact shape of the vertical density profile, and $\rho_\mathrm{sd}$ is the density of stars plus dark matter in the midplane. For convenience we adopt the simple assumptions mentioned in \citet{ostriker_regulation_2010}, 
\begin{equation}
\label{eq:rhosd}
\rho_\mathrm{sd} \approx \frac{v_\mathrm{circ}^2/r^2}{4\pi G} + \frac{\pi G \Sigma_*^2}{2 \sigma_*^2},
\end{equation}
where the first term comes from assuming that $v_\mathrm{circ}$ from dark matter is constant with radius, and the second comes from taking the stars to be locally isothermal with little modification in their vertical profile from the self-gravity of the gas. Caution is warranted in adopting both of these in high-z gas-rich disks, given that more explicit mass models find that the disks are largely baryon-dominated \citep{genzel_strongly_2017, price_rotation_2021}, and the gas fractions are high enough \citep{tacchella_sins_2015} that the gas will likely have some effect on the stellar scaleheight. However, these same effects that make Equation \ref{eq:rhosd}'s estimate of $\rho_\mathrm{sd}$ slightly worse also make $\rho_\mathrm{sd}$ less relevant in calculating $H$ (see Equation \ref{eq:H}).

Next we adopt the \citet{krumholz_star_2012} star formation law, in which the star formation rate surface density is proportional to the surface density of molecular gas, $f_{\mathrm{H}_2} \Sigma$, times an efficiency factor $\epsilon_\mathrm{ff} \approx 0.01$ \citep{krumholz_slow_2007, krumholz_star_2012}, divided by an estimate of the local effective freefall time.
\begin{equation}
    \dot{\Sigma}_\mathrm{SF} = \epsilon_\mathrm{ff} f_{\mathrm{H}_2} \Sigma / t_\mathrm{SF},
\end{equation}
The star formation timescale is the shorter of the following two timescales,
\begin{equation}
    t_\mathrm{SF}=\begin{cases}
    t_\mathrm{GMC} = \frac{\pi^{1/4}}{\sqrt{8}} \frac{\sigma}{G (\Sigma_0^3 \Sigma)^{1/4} }, & \mathrm{if}\ t_\mathrm{GMC}<t_Q\\ 
    t_\mathrm{Q} = \frac1\Omega\sqrt{\frac{3\pi^2 Q^2}{32(\beta+1)\phi_P}}, &  \mathrm{if}\ t_\mathrm{GMC}>t_Q .
    \end{cases},
\end{equation}
where $\phi_P$ is a constant that we set to 3, and $\Sigma_0$ is $85\ M_\odot\ \mathrm{pc}^{-2}$. The molecular gas fraction $f_{\mathrm{H}_2}$ is set according to the \citet{krumholz_star_2013} model, which is metallicity-dependent, but in the regime of interest where $Q\le Q_\mathrm{crit}$, usually $f_{\mathrm{H}_2}\approx 1$. Finally, the accretion rate is set following \citet{bouche_impact_2010}, who specify simple powerlaw accretion rates as a function of halo mass and redshift. Halo masses are estimated by inverting the median \citet{moster_galactic_2013} stellar mass-halo mass relation, and the radial dependence of the accretion rate surface is taken to follow a simple exponential with a scale length of 0.1 times the Virial radius \citep{forbes_radially_2019}.

\subsection{Gravitational Instability}
In equilibrium the source terms discussed in the previous subsection will be cancelled by the effects of gravitational instability, which simultaneously induces radial transport and heats the disk. One possibility following stellar N-body simulations \citep{sellwood_spiral_1984, carlberg_dynamical_1985} is to take
\begin{equation}
    \frac{dQ}{dt}\Big|_\mathrm{GI} = \frac{\mathcal{G}(Q)}{t_\mathrm{orb}},
\end{equation}
where $\mathcal{G}$ is some dimensionless function, and $t_\mathrm{orb} = 2\pi r/v_\mathrm{circ}$ is the local orbital time. \citet{sellwood_spiral_1984} found $\mathcal{G} = (1/4)(Q_\mathrm{crit}-Q)$ when $Q<Q_\mathrm{crit}$ and 0 otherwise. This illustrates some of the properties we might expect of $\mathcal{G}$, namely that $\mathcal{G}(Q=Q_\mathrm{crit})\approx 0$, then monotonically increases as $Q$ decreases. Arguably $\mathcal{G}(Q)$ must also depend only on $Q$ and not on any other parameters, since $Q$ should totally encapsulate the degree of instability. In general this reasoning is not totally certain because the physics of the non-linear evolution of the stability is not just a function of $Q$, but also the physics of feedback \citep{goldbaum_mass_2015, goldbaum_mass_2016, behrendt_clusters_2016}.

An alternative starting point is to assume that the GI term can be modeled as a viscosity \citep{balbus_dynamical_1999, gammie_nonlinear_2001, lodato_testing_2005, krumholz_dynamics_2010}, in which case we can write out the corresponding rate of change in $Q$ explicitly, equivalently as a function of viscosity, a \citet{shakura_reprint_1973} $\alpha$ parameter, or a vertically integrated torque $\mathcal{T}$. These quantities are related via
\begin{equation}
    \nu = \alpha \sigma^2/\Omega
\end{equation}
which serves as a definition of $\alpha$, and
\begin{equation}
    \mathcal{T} = 2\pi r \Sigma v_\mathrm{circ} (\beta -1) \nu
\end{equation}
\citep[see][for instance]{shu_physics_1992}. The change in $Q$ due to a local torque is then
\begin{equation}
\label{eq:torque}
    \frac{dQ}{dt}\Big|_\mathrm{GI} \approx  \frac{\partial Q}{\partial \sigma} \frac{(\beta-1) v_\mathrm{circ} \mathcal{T}}{6\pi r^3 \Sigma \sigma} = \frac{\partial Q}{\partial \sigma} \frac13 (\beta-1)^2  \alpha \sigma \Omega,
\end{equation}
Equation \ref{eq:torque} is an approximation because terms related to the advection of energy through the disk that contribute to $\partial \sigma / \partial t$ have been dropped, as has a term related to the accumulation (or loss) of mass because of differential mass flow, proportional to $\partial^2 \mathcal{T}/\partial r^2$. The two energy terms, proportional to $\partial \mathcal{T}/\partial r^2$ and $\mathcal{T} \partial \sigma/\partial r$, as well as the mass flow term, are all zero in the equilibrium solutions found by \citet{krumholz_dynamics_2010} and \citet{krumholz_unified_2018}, and regardless are non-local in the sense that they depend on the structure of the disk instead of conditions at one radius. The \citet{krumholz_dynamics_2010} solution was obtained by solving the radial differential equation for $\mathcal{T}$ under the assumption that $Q=Q_\mathrm{crit}$ exactly \citep[see also][in the context of protoplanetary disks]{rafikov_properties_2009, rice_stability_2011, rafikov_viscosity_2015}. In other words, the torque/viscosity/$\alpha$ is set based on global conditions in the disk in order to maintain $Q=Q_\mathrm{crit}$.

One possibility at this stage is to specify a particular $\mathcal{G}(Q)$, then solve for $\mathcal{T}$ everywhere in the disk simultaneously via a similar ordinary differential equation in $\mathcal{T}(r)$. This approach was taken by \citet{forbes_balance_2014} to compute the radial motions and heating of stars subject to spiral arms in a way that obeyed the \citet{sellwood_spiral_1984} result, and by \citet{krumholz_vader_2015} to illustrate numerically how a disk would evolve towards $Q=Q_\mathrm{crit}$. When the advection terms are dropped, no differential equation is needed and one need only specify an explicit form of $\mathcal{T}$ or $\alpha$ dependent on local conditions. In this case we would arrive at
\begin{equation}
\label{eq:gapprox}
    \mathcal{G} = t_\mathrm{orb} \frac{d Q}{dt}\Big|_\mathrm{GI} \approx  \frac{\partial Q}{\partial \sigma} \frac{2\pi}{3} (\beta-1)^2  \alpha \sigma \sim \frac{2\pi}{3} (\beta-1)^2 \alpha Q
\end{equation}
where the last approximation applies in the limit where $Q$ is set mostly by the instability of the gas. This form makes it clear that $\mathcal{G}$ plausibly really is just a function of $Q$ (and the approximately fixed rotation curve), provided that $\alpha$ itself is only a function of $Q$.

In the protoplanetary disk literature, a variety of $Q$-dependent values of $\alpha$ have been proposed, though all are fairly ad-hoc, essentially aiming to satisfy the expectations we had for $\mathcal{G}$, namely that it's close to zero at $Q=Q_\mathrm{crit}$ and rises monotonically as $Q$ decreases. Examples include
\begin{equation}
    \alpha_\mathrm{Zhu} = \exp{\left(-Q^4\right)}
\end{equation}
\citep{zhu_longterm_2010, zhu_longterm_2010a} and 
\begin{equation}
    \alpha_{LP} \propto \left(\left( \frac{Q_\mathrm{crit}}{Q} \right)^2 - 1\right)
\end{equation}
\citep{lin_formation_1987, lin_formation_1990}. Note that these prescriptions for $\alpha(Q)$ are separate from the prescriptions that arise from a calculation that assumes $Q=Q_\mathrm{crit}$ exactly, which also produce estimates of $\alpha$ \citep[e.g.][]{pringle_accretion_1981, gammie_nonlinear_2001, rafikov_properties_2009, krumholz_dynamics_2010}, since these values only apply at the critical value of $Q$ \citep{kratter_gravitational_2016}.  A compilation of $\alpha$ from simulations \citep{kratter_global_2008} provides a smooth transition in $\alpha$ as $Q$ goes from 2 to 1, with some additional dependence on the ratio of the disk mass to the stellar mass, but then assumes a saturation below $Q=1$ since the disk begins to fragment. There is presumably no recovering from this in a protoplanetary disk, but stellar feedback and the ability of stars as a large population to self-regulate their own value of $Q$ mean that galaxies can plausibly equilibrate in this regime. However, given the variety of $\alpha$ prescriptions available and their acknowledged ad-hoc nature, the next step is to compare to data.

\section{Comparison to Data}
\label{sec:data}

\begin{figure}
    \centering
    \includegraphics[width=0.9\textwidth]{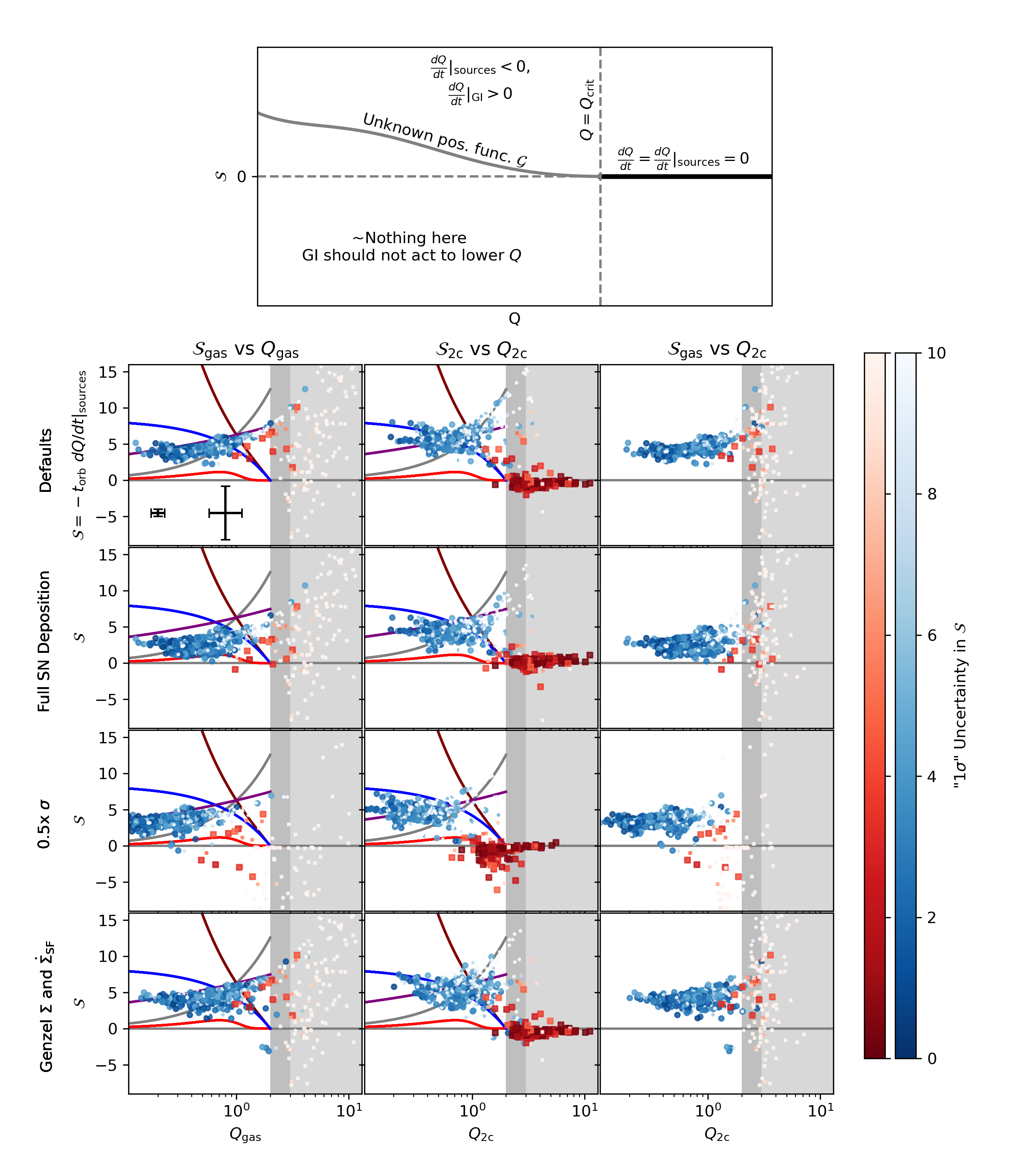}
    \caption{Data (forcing, $\mathcal{S}$, points) versus theory (GI response, $\mathcal{G}$, lines). Every panel shows some version of $\mathcal{S}=-t_\mathrm{orb} dQ/dt|_\mathrm{sources}$ as a function of some version of $Q$. The top panel shows a schematic of what we expect each panel to look like: gravitational instability has no effect for $Q>Q_\mathrm{crit}$, then acts to increase $Q$ at some unknown rate as a function of $Q$ below $Q_\mathrm{crit}$. In equilibrium, the source terms (estimated from the data) will exactly cancel this effect. The points show radius-by-radius estimates of $\mathcal{S}$, i.e. the negative rate of change of $Q$ per dynamical time owing to star formation, accretion, and turbulent dissipation. The red squares are from the DiskMass sample at $z\sim 0$, and the blue circles are from the SINS $z\sim 2$ sample. The lines show prescriptions for $\mathcal{G}=t_\mathrm{orb} dQ/dt|_\mathrm{GI}$, the rate of change of $Q$ per dynamical time owing to gravitational instability. From highest to lowest at $Q=2$, we show lines corresponding to constant $\alpha=3$ (gray), $\alpha \propto Q^{-0.75}$ (purple), \citet{lin_formation_1987} (maroon), $\alpha \propto (2/Q-1)$ (blue), and \citet{zhu_longterm_2010a} (red). Uncertainties shown by the colorbars are estimated by varying each point according to Table \ref{tab:perturbations} and taking the standard deviation of the perturbed values of $\mathcal{S}$. The range of $Q_\mathrm{crit}$ suggested by \citet{elmegreen_gravitational_2011}, from 2-3, is highlighted as a vertical gray bar in the plots, with higher values of $Q$ shown in a lighter gray. Typical reported statistical errors and estimated systematic errors for the SINS sample are shown as the errorbars in the top left panel.}
    \label{fig:genzel}
\end{figure}

\begin{table}[h!]
\centering
\begin{tabular}{c c c c} 
 \hline
 Ingredient for Estimating $\mathcal{S}$ & Perturbation for DiskMass Data & Perturbation for SINS Data  \\ [0.5ex] 
 \hline\hline
 $\Sigma$ & 0.1 dex & 0.3 dex  \\ 
 $\sigma$ & 0.08 dex times log-uniform from 0.5 to 1 & 0.15 dex  \\
$\Sigma_*$ & 0.18 dex & 0.3 dex  \\ 
 $\sigma_*$ & 0.1 dex times log-uniform from 1 to 2 & 0.48 dex  \\
 $v_\mathrm{circ}$ & 0.04 dex & 0.15 dex  \\
  $\beta=d\ln v_\mathrm{circ}/d\ln r$ & 0.1 & 0.1  \\
 $\dot{\Sigma}_\mathrm{SF}$ & 0.3 dex & 0.3 dex  \\
 $Z/Z_\odot$ & 0.3 dex & 0.3 dex  \\
$\rho_\mathrm{dm}$ & 0.3 dex & 1 dex  \\
$\langle p_*/m_* \rangle$ & 0.11 dex & 0.3 dex \\
 $\eta$ & 0.3 dex & 1 dex  \\ [1ex] 
 \hline
\end{tabular}
\caption{Perturbations to the observational data to estimate the size of each point's systematic uncertainty. Most perturbations are multiplicative with the exception of those for $\beta$. $\beta$ is additionally clipped to be between -0.5 and 1.}
\label{tab:perturbations}
\end{table}

\begin{table}[h!]
\centering
\begin{tabular}{c c c } 
 \hline
 $\alpha = 3\mathcal{G}/(2\pi (\beta-1)^2 Q)$ & Reference & Color in Figure \ref{fig:genzel}  \\ [0.5ex] 
 \hline\hline
 3 & - & Gray  \\ 
 $3/Q^{0.75}$ & This work & Purple  \\
 $2^2/Q^2 - 1$ & \citet{lin_formation_1987}, \citet{lin_formation_1990} & Maroon  \\
 2(2/Q - 1) & This work & Blue  \\
 $\exp{\left(-Q^4\right)}$ & \citet{zhu_longterm_2010a}, \citet{zhu_longterm_2010} & Red  \\ [1ex] 
 \hline
\end{tabular}
\caption{Values of $\alpha$ viscosity and the color of their corresponding lines in Figure \ref{fig:genzel}}
\label{tab:alphas}
\end{table}

We now turn to two published datasets where we can estimate $Q$ and the magnitude of each source term, and where we know there are cases where $Q$ is observed to be below plausible values of $Q_\mathrm{crit}$. These are the SINS sample of galaxies \citep{forsterschreiber_sins_2009}, including fits to the dynamics and mass profiles of 19 galaxies at $z \sim 2$ \citep{genzel_sins_2014, tacchella_sins_2015}, and the DiskMass sample \citep{bershady_diskmass_2010, martinsson_diskmass_2013, westfall_diskmass_2014} of 30 local nearly face-on galaxies. Our goal is to estimate the value of $dQ/dt|_\mathrm{sources}$ for each measured annulus in these two galaxy samples, and check whether they can be cancelled out by $dQ/dt|_\mathrm{GI}$ below $Q_\mathrm{crit}$.  

The SINS data are largely based on H$\alpha$ line emission whose flux, centroid, and width can be measured. From this a dynamical model is constructed assuming a velocity dispersion that is constant in radius, and assuming that the H$\alpha$ emission itself comes from star formation, from which a surface density of gas can be inferred via inversion of the Kennicutt-Schmidt relation. We enumerate the published values of $v_\phi(r)$, $\Sigma(r)$, and $Q_\mathrm{gas}(r)$, the latter two of which also include errorbars based on uncertainties in the H$\alpha$ flux. Each galaxy contains $\sim 20$ radial bins along the main axis of the galaxy for $\Sigma$ and $Q_\mathrm{gas}$. We take each of these points and plug them in to Equation \ref{eq:sources} assuming $\eta=1$, $Z=0.5 Z_\odot$, $\langle p_*/m_*\rangle$ from \citet{orr_bursting_2022}, $\epsilon_\mathrm{accr} = 0.1$, and $\dot{\Sigma}_\mathrm{accr}$ assuming the exponential scalelength of accretion is 10\% of the Virial radius \citep{forbes_radially_2019}, where the Virial radius and halo mass are estimated by inverting the median relation between $M_*$ and $M_h$ from \citet{moster_galactic_2013}. The normalization of $\dot{\Sigma}_\mathrm{accr}$ uses the simple powerlaw scaling with halo mass and $(1+z)$ from \citet{bouche_impact_2010} with an efficiency of 10\%. The stellar surface densities are taken from the single- or two-component surface density models of \citet{tacchella_sins_2015} derived from H-band photometry. We adopt the two-component model where available, and the one-component model otherwise. If neither model is available for a given galaxy, it is excluded from the sample. We also check that the adopted values of $Q_\mathrm{gas}$, $\Sigma$, and $v_\mathrm{circ}$ yield a constant $\sigma$, and exclude any annuli where this is not the case owing to digitization errors. Since 
the quoted $\Sigma$ in these observations is inferred from the H$\alpha$ luminosity by assuming a particular redshift-dependent molecular gas depletion time, but we use the \citet{krumholz_metallicitydependent_2012} star formation law (see Section \ref{sec:sources}), we use the quoted $\Sigma$ values to find the original estimated star formation rate surface density, then use the \citet{krumholz_metallicitydependent_2012} star formation law to infer $\Sigma$. In other words, we keep the estimate of the star formation rate surface density unchanged, but infer a different underlying gas column density. The effect of this change is among the sources of systematic error that we explore.

Finally, in order to estimate a two-component $Q$, which we will call $Q_\mathrm{2c}$, we need to assume something about the velocity dispersion of the stars, $\sigma_*$. We adopt a smooth interpolation between the case that $\sigma_*\approx\sigma$ as the result of recent star formation, and $Q_*=Q_{*,\mathrm{crit}}=2$, from the action of spiral arm heating. In particular, 
\begin{equation}
\frac{\sigma_*}{\sigma} =\frac{\mathcal{I}_\mathrm{spiral}\sigma_{*,\mathrm{crit}} + \sqrt{\mathcal{I}_\mathrm{spiral}\sigma_{*,\mathrm{crit}}^2 + (\phi^2 + 2\phi\mathcal{I}_\mathrm{spiral}) \sigma^2} }{ (\phi + 2 \mathcal{I}_\mathrm{spiral}) \sigma}
\end{equation}
Here $\phi=f_R\gamma\ \mathrm{sSFR}\ t_\mathrm{orb}$, $\sigma_{*,\mathrm{crit}} = Q_{*,\mathrm{crit}} \pi G \Sigma_*/\kappa$, and $\mathcal{I}_\mathrm{spiral}=1$ when $\sigma < \sigma_{*,\mathrm{crit}}$ and zero otherwise. This formula is derived by assuming that $\partial \sigma_*/\partial t = 0$. In future work we will validate and refine this estimate, but for now we adopt it and keep this in mind as a source of systematic uncertainty for our analysis of the SINS galaxies.

In the DiskMass sample, we include all radii that have estimated values of $\sigma_z$, namely the stellar velocity dispersion in the direction perpendicular to the plane of the disk presented in \citet{martinsson_diskmass_2013a}, and line-of-sight velocity dispersions of [OIII]$\lambda5007\rm{\AA}$ as presented in \citet{martinsson_diskmass_2013}. For each annulus, we also incorporate the estimated HI and $\mathrm{H}_2$ surface densities from the Westerbork survey and Spitzer 24 $\mu$m photometry respectively \citep{bershady_diskmass_2010}. We estimate the local star formation rate surface density as $\Sigma_{\mathrm{H}_2}/(2\ \mathrm{Gyr})$ \citep{bigiel_constant_2011}. Stellar mass surface densities are estimated based on K-band photometry, and circular velocities are adopted from the dynamical modelling (Case Ia) of \citet{martinsson_diskmass_2013a}. The DiskMass survey corrects their observed line of sight stellar velocity dispersions to estimate $\sigma_z$, but typically the radial component of the stellar velocity dispersion is larger by up to a factor of two depending on the physical mechanism by which the stars were heated \citep[e.g.][]{lacey_influence_1984}, and it is the radial component of the velocity dispersion that enters $Q_*$. This is another source of systematic uncertainty.

When estimating $Q_\mathrm{2c}$ within the DiskMass collaboration \citep{westfall_diskmass_2014}, the authors take the observed ionized gas velocity dispersions and divide by two to estimate the cold gas velocity dispersion \citep{andersen_photometric_2006, caldu-primo_highdispersion_2013}. This difference in velocity dispersion also arises in simulations with resolved multi-phase ISMs \citep{jeffreson_scale_2022}. In our treatment where we take the gas to be a single component with a single velocity dispersion, it is not obvious which velocity dispersion to adopt in general. For the DiskMass sample, usually the molecular gas is subdominant in terms of mass, so the velocity dispersion of the disk may be better-represented by the velocity dispersion of the warmer component. \citet{girard_systematic_2021} argue that for $z\sim 2$ galaxy analogues observed at $z \sim 0$, there is a factor of $>2$ offset between molecular and ionized gas velocity dispersions. This is more serious since the $z\sim 2$ galaxies are dominated by molecular gas. On the other hand \citet{ubler_kinematics_2021} found in an analysis of TNG50 data that analyzing the simulation in the same way as the observations yields an offset that varies from galaxy to galaxy, with the observation-like pipeline yielding higher velocity dispersions, though often just a few km s$^{-1}$, but occasionally up to a factor of two. By default we therefore do not adjust the ionized velocity dispersions, but acknowledge that the velocity dispersions may plausibly be lower by a factor of 2. Keep in mind that this plausible adjustment factor may be different between the two samples.

Having now collected all of the assumptions necessary to estimate $dQ/dt|_\mathrm{sources}$ for every annulus in these two samples, we now plot the results. Figure \ref{fig:genzel} shows 12 variations of $\mathcal{S}$ vs. $Q$, and a schematic representation of what we might expect this plot to look like a priori. We define $\mathcal{S} \equiv -t_\mathrm{orb} dQ/dt|_\mathrm{sources}$, which is the rate of change of $Q$ per local dynamical time. The minus sign allows us to directly compare $\mathcal{S}$ to possibilities for $\mathcal{G}$, the rate of change of $Q$ per dynamical time from the effects of gravitational instability, since in equilibrium $\mathcal{S} = \mathcal{G}$. Schematically, we expect that $\mathcal{S}=\mathcal{G}=0$ when $Q>Q_\mathrm{crit}$, and for $\mathcal{S}$ and $\mathcal{G}$ to trace out some positive function for $Q<Q_\mathrm{crit}$. Each column shows a different version combination of $\mathcal{S}$ and $Q$ -- the subscripts indicate which version of $Q$ is used to evaluate either $Q$ itself or in the case of $\mathcal{S}$, the partial derivatives of $Q$ in Equation \ref{eq:sources}. Note that to evaluate $\mathcal{S}_\mathrm{gas}$, the last three terms of Equation \ref{eq:sources} are all zero because $Q_\mathrm{gas}$ has no explicit dependence on the stellar component. Each row shows an alternative calculation of $\mathcal{S}$ based on a plausible change in our assumptions. The second row shows the effect of ignoring the \citet{orr_bursting_2022} model and simply using $\langle p_*/m_*\rangle = 3000\ \mathrm{km}\ \mathrm{s}^{-1}$, the third row shows a systematic reduction in observed values of $\sigma$ by a factor of 2, and the final row shows the adoption of the \citet{genzel_sins_2014} values of $\Sigma$ based on the observed H$\alpha$ flux for those galaxies (the DiskMass sample is unchanged between the first and fourth row). 

Each point in Figure \ref{fig:genzel} is shaded according to its estimated systematic error, with larger errorbars shown as lighter colors, and uncertainties greater than 5 also using a much smaller symbol size. The red squares show the DiskMass sample, and the blue circles show the SINS sample. The errorbars in the top left panel show the median statistical and systematic uncertainties for the SINS sample. For each point the systematic error is estimated by re-calculating $\mathcal{S}$ 10 times with perturbed values of the default input parameters and recording the standard deviation of the samples. A more principled approach would be to fit the data with an explicit model and representing the systematic offsets as parameters to be fit drawn from their own distributions per annulus and per galaxy, but short of that we are able to see which points are reliable in this space, and which are more uncertain. The adopted random perturbations are shown in Table \ref{tab:perturbations}. Remarkably much of the DiskMass sample, particularly the annuli with $Q_\mathrm{gas} \ga 3$, is subject to large uncertainties in $\mathcal{S}_\mathrm{gas}$. These arise largely from the modest uncertainty in the star formation rate, since in this regime we expect no reduction in $\langle p_*/m_* \rangle$, and the low velocity dispersions of the gas make it quite sensitive to the effects of feedback (see also Section \ref{sec:interpretation}). Meanwhile $\mathcal{S}_\mathrm{2c}$ for this sample is insensitive to these uncertainties because $Q_\mathrm{2c}$ is dominated by the stellar component.

For comparison to the data, we plot a number of curves for $\mathcal{G}$, namely plausible but ad-hoc descriptions of how gravitational instability will behave as a function of $Q$. The lines are identical in all 8 panels where they are shown in Figure \ref{fig:genzel}. Many of them arise from a particular formulation of $\alpha$, but notably there is not a strict one-to-one translation between $\alpha$ and the two-component version of $\mathcal{G}$, which we could call $\mathcal{G}_\mathrm{2c}$. That is, once the stellar component is included, an $\alpha$ viscosity that acts by heating and transporting only gas will have a different effect on $Q_\mathrm{2c}$ depending on a wide range of other parameters. Therefore when we show the same line in the middle panel as we showed in the left panel, $\alpha$ itself is not the same. Generally the same $\alpha$ would lead to lower values of $\mathcal{G}_\mathrm{2c}$ than $\mathcal{G}_\mathrm{gas}$. Recall from Equation \ref{eq:gapprox} that $\mathcal{G}_\mathrm{gas} = 2\pi(\beta-1)^2\alpha Q_\mathrm{gas}/3$, and this is how we translate the values of $\alpha$ in Table \ref{tab:alphas} to the lines shown in Figure \ref{fig:genzel}.

\section{Interpretation}
\label{sec:interpretation}

We are searching for the function $\mathcal{G}(Q)$ that balances out every other effect on $Q$ so that $dQ/dt \approx 0$, and which approaches zero for $Q \ga Q_\mathrm{crit}$, whose exact value is also not known. It is also not obvious what version of $Q$ to use, i.e. the gas-only version, or the multi-component version. This is why in Figure \ref{fig:genzel} we show both $\mathcal{S}_\mathrm{gas}$ vs $Q_\mathrm{gas}$ and $\mathcal{S}_\mathrm{2c}$ vs $Q_\mathrm{2c}$. This ambiguity arises because $Q$ and its various versions only specify the stability of the disk in the linear regime with no cooling, whereas per \citet{elmegreen_gravitational_2011}, realistic disks are never formally stable. We therefore examine the evidence for a purely gas instability-driven effect, and alternatively an instability in which the stars also participate.

In the gas-driven case, the relevant panels of Figure \ref{fig:genzel} are the first column, showing $\mathcal{S}_\mathrm{gas}$ as a function of $Q_\mathrm{gas}$. The data as a whole, largely dominated by the SINS data given the systematic uncertainties for disks in the DiskMass regime, show a pattern similar to what we might expect a priori. The data have $\mathcal{S}$ convincingly above zero until $Q_\mathrm{gas} \ga 3$, at which point the errorbars become too large to reliably discern much of a trend. When $Q_\mathrm{gas} \la 2$, the data are close to a line corresponding to $\alpha = 3/Q_\mathrm{gas}^{0.75}$.

In the two-component case, the data quite clearly show that $\mathcal{S} \approx 0$ for $Q_\mathrm{2c} \ga 2$, a regime well-populated in the DiskMass sample. The SINS data display a larger scatter than in the gas-driven case, although this may be driven by a combination of greater systematic uncertainties given the higher redshift and the fact that $\sigma_*$ is not measured but guessed in our analysis. Roughly speaking the data follow the blue line, corresponding to $\mathcal{G}_\mathrm{2c} = (4\pi/3) (\beta-1)^2 Q_\mathrm{2c} (2/Q_\mathrm{2c}-1)$. In this version we would also expect that $\mathcal{S}_\mathrm{gas}$ should be zero when the instability is inactive, i.e. when $Q_\mathrm{2c}>Q_\mathrm{crit}$. Showing this relationship in the third column of Figure \ref{fig:genzel}, we see that on balance $\mathcal{S}_\mathrm{gas}$ is above zero, though this hinges on only a handful of points with substantial errorbars. 

While these datasets do not immediately tell us exactly which version of $Q$ and $\mathcal{G}$ to use, each prescription is similar in the low-$Q$ regime, since in these samples low $Q_\mathrm{2c}$ only occurs when $Q_\mathrm{2c}$ is dominated by the gas component. Since this is the regime that originally motivated our investigation into how low $Q$ could go, it is worth examining $\mathcal{S}$ in this regime to better understand how exactly the data end up showing $\mathcal{S} \sim 5$.
In this limit, namely that $Q$ is set almost entirely by the gas,  $f_{\mathrm{H}_2} \approx 1$, and $t_\mathrm{SF}\approx t_Q$,
\begin{align}
\label{eq:gasdom}
    t_\mathrm{orb}\frac{dQ}{dt}\Big|_\mathrm{sources} \sim & 2\epsilon_\mathrm{ff}\sqrt{\frac{32(\beta+1)\phi_P}{3}}\left(f_R+\eta+\frac13\frac{\tilde{Q}_\mathrm{SN}}{Q}\right) \\ \nonumber &+\frac{\dot{\Sigma}_\mathrm{accr} t_\mathrm{orb}}{\Sigma}Q\left(\epsilon_\mathrm{accr} \frac{\tilde{Q}_\mathrm{orb}^2}{Q^2} -1 \right) \\ \nonumber
    & -\pi\sqrt{\frac{\beta+1}{2}}\left( 1 + \left(1+4 \zeta_d Q^2 \left( \frac{1}{(\beta+1)} + \frac{8\pi }{Q_*^2}\right) \right)^{1/2} \right) \left(1 - \frac{\tilde{Q}_\mathrm{therm}^2}{Q^2} \right)^{3/2}
\end{align}
To non-dimensionalize the equation, we have defined several $Q$-like quantities here, where in each case the velocity dispersion has been replaced with a different relevant velocity. In particular, $\tilde{Q}_\mathrm{orb} \equiv v_\mathrm{orb}\kappa/(\pi G \Sigma)$ and $\tilde{Q}_\mathrm{SN} \equiv \langle p_*/m_* \rangle \kappa/(\pi G \Sigma)$. In both cases, the ratio of these $\tilde{Q}$'s to $Q$ is often $\gg 1$, though given the large velocity dispersions in these galaxies $\tilde{Q}_\mathrm{orb}/Q$ may be less than 1 in galaxy centers. As in \citet{forbes_balance_2014}, we include a factor in the final term containing $\tilde{Q}_\mathrm{therm} \equiv \sigma_\mathrm{therm} \kappa/(\pi G\Sigma)$ to reduce the dissipation term to zero as $\sigma \rightarrow \sigma_\mathrm{therm}$, i.e. where the velocity dispersion arises purely from the thermal soundspeed of the Warm Neutral Medium.

\begin{figure}
    \centering
    \includegraphics[width=\textwidth]{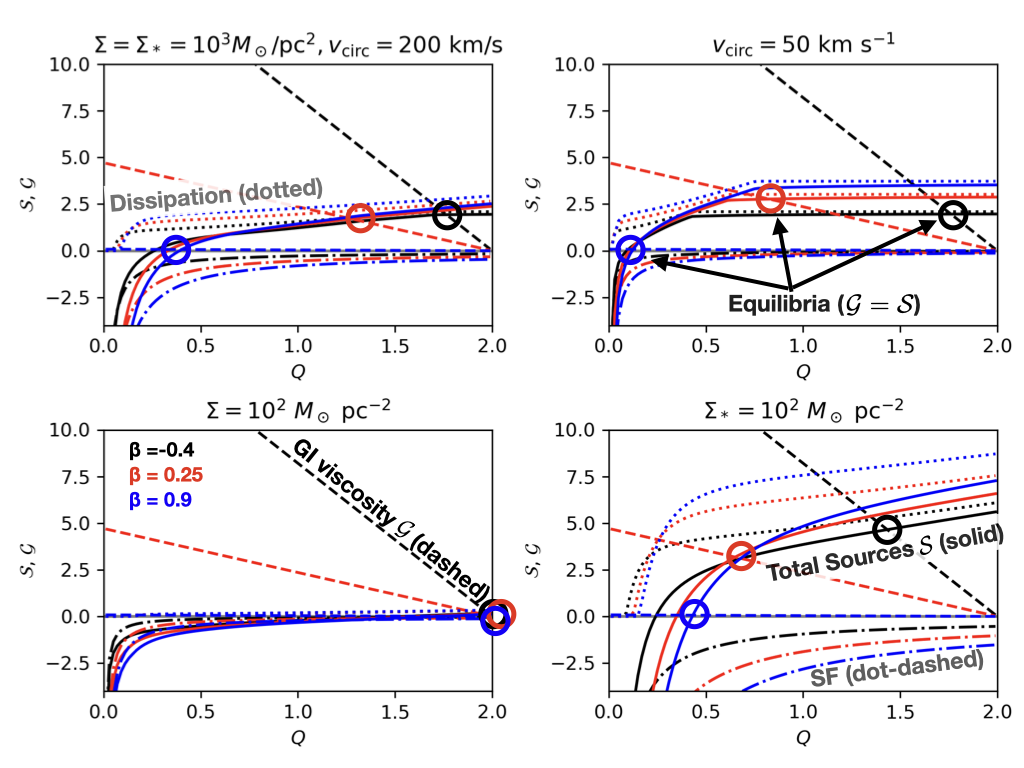}
    \caption{The dependence of star formation, turbulent dissipation, and gravitational instability on $Q$. The dashed lines showing the effect of gravitational instability as encapsulated by $\mathcal{G} = t_\mathrm{orb} dQ/dt|_\mathrm{GI}$ are the same in each panel, and vary quite dramatically as a function of $\beta$. We have adopted a particular form of $\mathcal{G}$ referred to in the text as the two-component case. In equilibrium, $\mathcal{G}=\mathcal{S}$, shown as the solid lines. These equilibria for each value of $\beta$ are highlighted as three circles in each panel. The total sources $\mathcal{S}$ are the sum of the effect of star formation (dot-dashed lines), which always acts to increase $Q$, and dissipation (dotted lines), which always act to decrease $Q$.}
    \label{fig:stability}
\end{figure}

In this limit, the loss term ranges from about 2-7, with a dependence on $Q$ that is at most linear under the right circumstances. Meanwhile star formation, both via the removal of gas and the injection of momentum, always opposes the loss term in a $Q$-dependent way, so that star formation becomes more effective at lower $Q$. In some models this is the end of the story -- $\epsilon_\mathrm{ff}$ changes such that these two terms exactly cancel at $Q=Q_\mathrm{crit}$ \citep[e.g.][]{faucher-giguere_feedbackregulated_2013}, but if $\epsilon_\mathrm{ff}$ is constant \citep{krumholz_slow_2007, krumholz_star_2012}, they will only cancel at one value of $Q$ depending on the gas fraction, the mass loading factor, and the momentum injection from SN feedback. The remaining term representing the effects of accretion may be positive or negative, since energy, but also mass, is added to the system. While in principle it appears that this term could dominate in the outer parts of disks where $\tilde{Q}_\mathrm{orb}/Q \gg 1$, $t_\mathrm{orb}$ is long, and $\Sigma$ is low, in that regime $\sigma$ is usually already at the thermal sound speed when $Q=Q_\mathrm{crit}$, meaning that in practice this term does not appear to matter very often.

The gas-dominated limit shown in Equation \ref{eq:gasdom} is useful in understanding Figure \ref{fig:stability} where we show how the source terms tend to vary as a function of $Q$ for a few different parameter combinations, along with the other side of the equation, $\mathcal{G} = t_\mathrm{orb} (dQ/dt)|_\mathrm{GI}$. For the latter we have set $\mathcal{G} = (4\pi/3) (\beta-1)^2 Q (2/Q -1 )$, which is a reasonable description of the data. The different colors correspond to different values of the rotation curve slope $\beta$: black, red, and blue are respectively $-0.45$, $0.25$, and $0.95$. These values have a huge effect on $\mathcal{G}$, but only modest effects on the source terms. This is because the viscosity is much less effective when there is little inherent shear in the disk (as $\beta \rightarrow 1$ and the disk approaches solid body rotation). 

Generally the source terms increase with increasing $Q$ as the influence of star formation wanes and the loss term becomes large. The source term on its own can be zero when star formation is important, and the loss term can be zero when $Q$ is so low that, at fixed values of $v_\mathrm{circ}$, $r$, $\beta$, and $\Sigma$, the velocity dispersion reaches the thermal soundspeed of the disk (taken here to be 8 km s$^{-1}$). For low values of $\beta$ where the disk experiences substantial shear, $\mathcal{G}$ is large enough to intersect the source terms at $Q$ only modestly less than $Q_\mathrm{crit}$, whereas at high values of $\beta$, it takes much lower values of $Q$ before $\mathcal{G}$ can counteract the source terms for this choice of viscosity. Note also that the intersections of the source and GI terms are almost always stable, since for example to the left of the equilibrium point, $\mathcal{S}<\mathcal{G}$, so $dQ/dt>0$, which would move the disk back towards the equilibrium point.

With this understanding of the gas-dominated regime, it is clear that the SINS data simply represent a set of galaxies where the turbulent dissipation rate is large enough that it cannot be balanced by the effects of star formation, even when stellar feedback is able to add a momentum per unit mass of 3000 km s$^{-1}$ (see second row of Figure \ref{fig:genzel}) or when the velocity dispersions are substantially over-estimated (see the third row of Figure \ref{fig:genzel}). This is very similar to previous results that have shown that high velocity dispersions cannot be maintained by supernovae alone \citep[e.g.][]{joung_dependence_2009,brucy_largescale_2020}, but here we show that the effect of gravitational instability is likely not so strong that $Q \rightarrow Q_\mathrm{crit}$. Instead, $Q$ remains in equilibrium at values far less than 1.

\section{Implications}
\label{sec:implications}
\begin{figure}
    \centering
    \includegraphics[width=\textwidth]{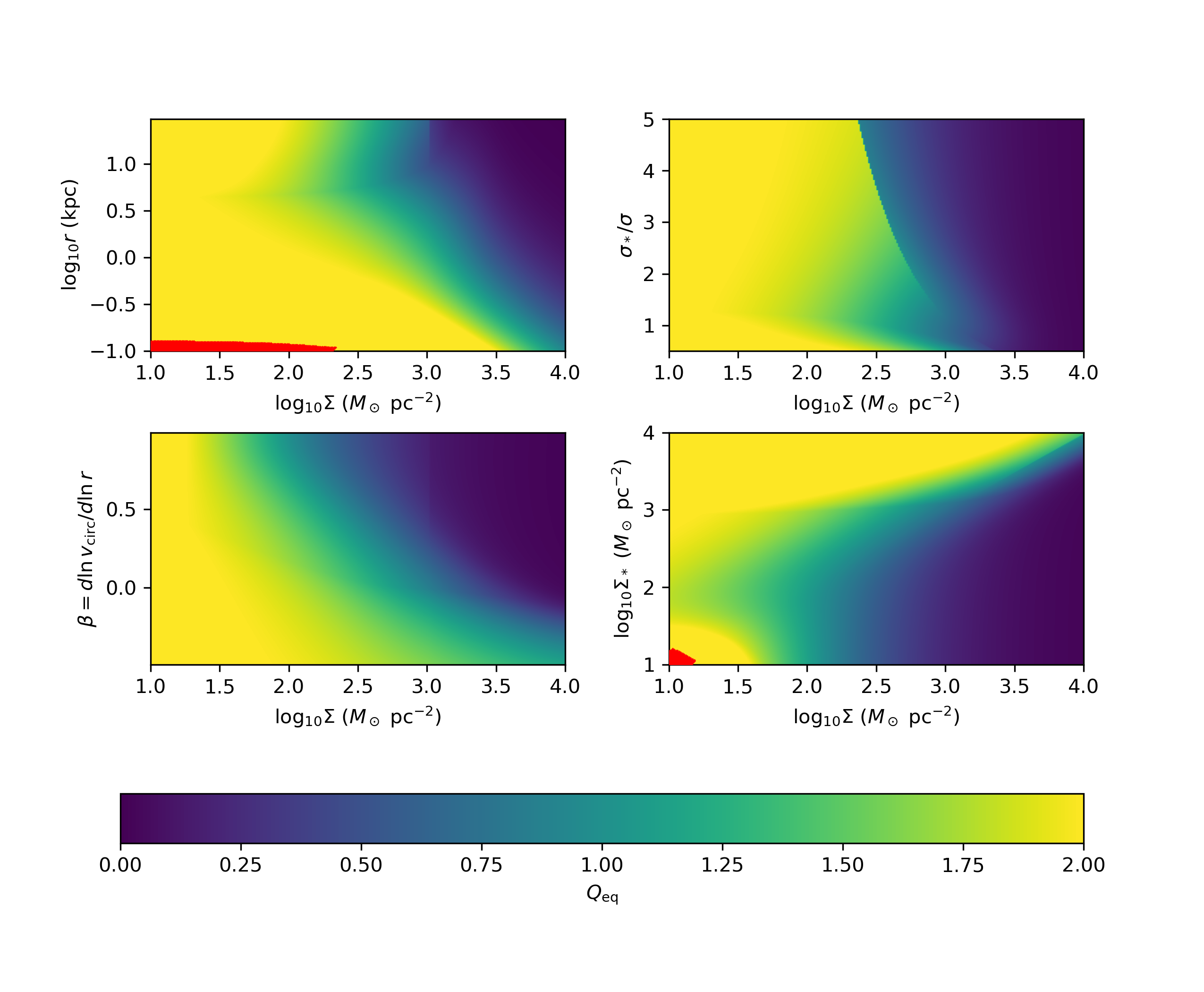}
    \caption{Equilibrium values of $Q_\mathrm{2c}$. Each panel shows a slice through parameter space. If a parameter is not being varied, it is set as follows: $\beta=0$, $v_\mathrm{circ} = 200\ \mathrm{km}\ \mathrm{s}^{-1}$, $r=5\ \mathrm{kpc}$, $\Sigma = \Sigma_*= 1000 M_\odot\ \mathrm{pc}^{-2}$. The red region indicates that $Q$ cannot fall below $Q_\mathrm{crit}$ because $\sigma$ is too low. Substantial parts of parameter space admit equilibrium values of $Q$ well below $Q_\mathrm{crit}$, particularly for high values of $\Sigma$ and $\beta$.}
    \label{fig:Qeq}
\end{figure}

By adopting a value of $\mathcal{G}(Q)$ that balances out the observed source terms, encapsulated as $\mathcal{S}$, we can check how varying the conditions in the disk will change $Q_\mathrm{eq}$, the value of $Q$ where $\mathcal{S}=\mathcal{G}$. We compute this solution numerically for $Q_\mathrm{2c}$, adopting Equation \ref{eq:sources} for the sources and $\mathcal{G} = (4\pi/3) (\beta-1)^2 Q (2/Q -1 )$ for the effect of gravitational instability. We then fix all but two values of $r$, $\sigma_*/\sigma$, $\Sigma$, $\Sigma_*$, $v_\mathrm{circ}$, and $\beta$, and plot the resulting equilibrium $Q_\mathrm{2c}$ in Figure \ref{fig:Qeq}. The fixed values are $r=5\ \mathrm{kpc}$, $\sigma_*/\sigma=1.1$, $\Sigma_*=10^3 M_\odot\ \mathrm{pc}^{-2}$, $v_\mathrm{circ}=200\ \mathrm{km}\ \mathrm{s}^{-1}$, and $\beta=0.1$. Large radii, small $v_\mathrm{circ}$, large $\Sigma_*$, or large $\beta$, will all, so long as $\Sigma \ga 10 M_\odot\ \mathrm{pc}^{-2}$, produce $Q_\mathrm{eq}$ noticeably lower than the assumed $Q_\mathrm{crit}=2$. The hatched region in red indicates that in that part of parameter space, $\sigma$ has reached the thermal sound speed at $Q=Q_\mathrm{crit}$, so $Q$ cannot fall below $Q_\mathrm{crit}$ there (and indeed these regions are surrounded by regions of $Q_\mathrm{eq} \approx Q_\mathrm{crit}$). 

The patterns in Figure \ref{fig:Qeq} can largely be understood by referring back to Figure \ref{fig:stability} and Equation \ref{eq:gasdom}. The simplest to understand is $\beta$, where increasing values decrease the shear in the disk, leading to weaker effects on $dQ/dt$ for a given value of $\alpha$. Interestingly this behavior with $\beta$ suggests that the phenomenon we are studying here, namely equilibrium values of $Q$ well below $Q_\mathrm{crit}$, is much harder to generate in protoplanetary disks than galactic disks, since in the former shear is always strong given the Keplerian rotation curves of $\beta=-1/2$. Variation in $r$ is a stand-in for variation in $\Omega$ and hence $\kappa$, which scale every version of $Q$ or $\tilde{Q}$ by the same factor, essentially stretching or narrowing many of the terms shown in Figure \ref{fig:stability} along the $x$-axis. Higher values of $\sigma_*/\sigma$ also tend to lower $Q_\mathrm{eq}$, sometimes quite sharply, as $Q_*$ becomes less important in determining $Q_\mathrm{2c}$. Essentially once $Q_*$ is important in setting $Q_\mathrm{2c}$, the partial derivatives of $Q$ with respect to $\Sigma$ and $\sigma$ become much smaller, and so the dramatic effects of star formation, feedback, and turbulent dissipation, are much weaker (see Equation \ref{eq:sources}). A similar effect is visible in the lower right panel showing variation in $\Sigma_*$.

While it is straightforward to compute $Q_\mathrm{eq}$ numerically once we adopt a particular function to represent $\mathcal{G}$ and have chosen a particular value of $Q_\mathrm{crit}$, it bears repeating that the data we have compiled here do not make it obvious precisely what $Q_\mathrm{crit}$ is, nor the value of $\mathcal{G}$ near $Q_\mathrm{crit}$, nor even which version of $Q$ to use. This dataset may be augmented by other surveys and by simulations, and the analysis may be improved by casting it as a hierarchical Bayesian inference problem. 

The ambiguities discussed here lead to uncertainty in how to implement our results here into simplified disk models that rely on $Q_\mathrm{crit}$ in some way \citep[e.g.][]{krumholz_dynamics_2010, forbes_evolving_2012, genel_effect_2012, cacciato_evolution_2012, forbes_balance_2014, porter_understanding_2014, krumholz_vader_2015, stevens_building_2016, krumholz_turbulence_2016, krumholz_unified_2018, forbes_radially_2019}. Clearly imposing that $Q \ge Q_\mathrm{crit}$ everywhere at all times is not adequate. However, the two different models for gravitational instability-induced changes in $Q$, namely the gas-only version and the multi-component version, may imply somewhat different prescriptions in simplified models. The gas-only version is likely the simplest, since there there is a direct one-to-one relationship between $\mathcal{G}$ and $\alpha$, so one could simply adopt $\alpha_\mathrm{GI} \approx 3/Q_\mathrm{gas}^{0.75}$ when $Q_\mathrm{gas}\la 2.5$ and $\approx 0$ otherwise. This $\alpha$ would have a local effect on $d\sigma/dt$ and $\dot{M}$, the rate of mass flow through the disk. The multi-component version where $\mathcal{G}_\mathrm{2c} = (4\pi/3)(\beta-1)^2 Q_\mathrm{2c}(2/Q_\mathrm{2c} - 1)$ is more ambiguous because a single value of $\alpha$ leads to a plethora of values of $\mathcal{G}_\mathrm{2c}$ depending on local conditions, and in general there is no guarantee that any value of $\alpha$ will be large enough to generate the required $\mathcal{G}_\mathrm{2c}$ when the stellar component is important in setting $Q_\mathrm{2c}$. Essentially because stars may participate in the instability in this formulation, they too may be altered by $\mathcal{G}$. The flatness of the data in the third column of Figure \ref{fig:genzel} suggests that one way to implement this may be to set $\alpha_\mathrm{GI} \propto 1/Q_\mathrm{2c}$ when $Q_\mathrm{2c}<Q_\mathrm{crit}$ and 0 otherwise, then calculating the $\mathcal{G}_* \equiv t_\mathrm{orb} dQ_*/dt|_\mathrm{GI}$ necessary to reach the given $\mathcal{G}_\mathrm{2c}$. Note that in general $\mathcal{G}_\mathrm{2c} \ne \mathcal{G}_\mathrm{gas} + \mathcal{G}_*$ because
\begin{equation}
\mathcal{G}_\mathrm{2c} = t_\mathrm{orb} \left( \frac{\partial Q_\mathrm{2c}}{\partial Q_\mathrm{gas}} \frac{\partial Q_\mathrm{gas}}{\partial t} + \frac{\partial Q_\mathrm{2c}}{\partial Q_*} \frac{\partial Q_*}{\partial t} +\frac{\partial Q_\mathrm{2c}}{\partial s} \frac{\partial s}{\partial t}\right)\Big|_\mathrm{GI},
\end{equation}
where $s$ is the ratio $\sigma/\sigma_*$. While backing out $\mathcal{G}_*$ from this equation is slightly more arduous than just subtracting off $\mathcal{G}_\mathrm{gas}$, it is straightforward to compute. Moreover this multi-component formulation of the effect of gravitational instability on $Q$ is more appealing theoretically simply because the stability of a multi-component disk definitely depends on both components in the linear regime. 

\section{Conclusion}
\label{sec:conclusion}

We attempt to reconcile the common theoretical assumption that $Q\ge Q_\mathrm{crit}$ with observations of $z\sim 2$ galaxies that show $Q \ll 1$ almost ubiquitously. This can be accomplished by relaxing this rigid theoretical assumption slightly and setting the viscosity according to a $Q$-dependent $\alpha$. Despite substantial observational uncertainties, a simple analytic form $\alpha\approx 3/Q_\mathrm{gas}^{0.75}$ for a gas-only version of the instability, or $dQ/dt|_\mathrm{GI} = (4\pi/3)(\beta-1)^2 (2 - Q_\mathrm{2c} ) t_\mathrm{orb}^{-1}$ for the multi-component version, would cancel out the rate of change of $Q_\mathrm{2c}$ due to other physical effects. Of these physical effects, the most important is the high rate of turbulent dissipation, tempered by the effects of star formation, implied by the high surface densities and velocity dispersions of the $z\sim 2$ galaxies. This provides a simple explanation for the low observed values of $Q$, and a path forward for improving the implementation of the effects of gravitational instability in simplified models.

\begin{acknowledgments}
We thank Rachel Somerville, Shy Genel, Mark Krumholz, Doug Lin, Chris Hayward, and Julianne Dalcanton, for helpful conversations. JCF is supported by a Flatiron Research Fellowship through the Flatiron Institute, a division of the Simons Foundation.
\end{acknowledgments}

\software{numpy \citep{vanderwalt_numpy_2011,harris_array_2020},
          matplotlib \citep{hunter_matplotlib_2007},
          scipy \citep{virtanen_scipy_2020}
          }

\bibliography{tngAccretion}{}
\bibliographystyle{aasjournal}

\end{document}